\documentclass[12pt,preprint2]{aastex}
\usepackage{url}

\shorttitle{Galaxy handedness asymmetry}
\shortauthors{Lior Shamir}

\begin{document}


\title{Asymmetry between galaxies with clockwise handedness and counterclockwise handedness}


\author{Lior Shamir\altaffilmark{1}}
\affil{Department of Computer Science, Lawrence Technological University, Southfield, MI, 48075}
\email{lshamir@mtu.edu}


\altaffiltext{1}{Lawrence Technological University, Southfield, MI, 48075}


\begin{abstract}
While it is clear that spiral galaxies can have different handedness, galaxies with clockwise patterns are assumed to be symmetric in all of their other characteristics to galaxies with counterclockwise patterns. Here we use data from SDSS DR7 to show that photometric data can distinguish between clockwise and counterclockwise galaxies. Pattern recognition algorithms trained and tested using the photometric data of a clean manually crafted dataset of 13,440 spiral galaxies with $z<0.25$ can predict the handedness of a spiral galaxy in $\sim$64\% of the cases, significantly higher than mere chance accuracy of 50\% ($P<10^{-5}$). Experiments with a different dataset of 10,281 automatically classified galaxies showed similar results of $\sim$65\% classification accuracy, suggesting that the observed asymmetry is consistent also in datasets annotated in a fully automatic process, and without human intervention. That shows that the photometric data collected by SDSS is sensitive to the handedness of the galaxy. 
Also, analysis of the number of galaxies classified as clockwise and counterclockwise by crowdsourcing 
shows that manual classification between spiral and elliptical galaxies can be affected by the handedness of the galaxy, and therefore galaxy morphology analyzed by citizen science campaigns might be biased by the galaxy handedness. Code and data used in the experiment are publicly available, and the experiment can be easily replicated.
\end{abstract}


\keywords{Galaxy: general -- galaxies: photometry -- galaxies: spiral}



\section{Introduction}
\label{introduction}


A highly noticeable morphological property of a spiral galaxy is its handedness. Spiral galaxies can be broadly separated into galaxies that seem to an Earth-based observer to have clockwise patterns, and galaxies that seem to have counterclockwise patterns. Since clockwise galaxies are expected to be symmetric to counterclockwise galaxies, this morphological difference is not expected to be reflected by other physical characteristics. The symmetry is also expected because the handedness of a galaxy is merely a matter of the location of the observer, and a galaxy that would seem to rotate clockwise to an Earth-based observer might seem to rotate counterclockwise to an observer placed elsewhere in the universe.

Some evidence show that the distribution of clockwise and counterclockwise spiral galaxies changes between different RA ranges, and may therefore violate the cosmological assumption of isotropy \citep{longo2011detection,shamir2012handedness}. Other studies showed some mild photometric differences between clockwise and counterclockwise galaxies \citep{shamir2013color}. However, unlike stars, galaxies cannot be considered a one-parameter family \citep{djorgovski1987fundamental}, and therefore a galaxy can be described by a set of multiple physical measurements \citep{brosche1973manifold,djorgovski1987fundamental}. 

When testing a large number of different measurements, the probability that a certain test exhibits a difference by mere chance increases as the total number of measurements gets higher. For instance, assuming that a single hypothesis can be considered statistically significant if the probability of false positive is smaller than 0.05, when testing multiple different hypotheses the probability that one of them exhibits a difference with statistical significance of P$<$0.05 is clearly higher than 0.05, and increases as the number of hypotheses being tested gets higher. Therefore, when multiple different hypotheses are being tested, the threshold of 0.05 must be corrected to avoid false positives.

A mature method to avoid false positives when testing a large number of hypotheses is the Bonferroni correction \citep{goeman2014multiple}, which provides the threshold of statistical significance that each specific hypothesis needs to meet when tested as part of an experiment that involves multiple hypotheses \citep{goeman2014multiple}. While the Bonferroni correction reduces the possibility of false positives, applying it makes it more difficult to identify statistically significant differences between clockwise and counterclockwise galaxies when comparing a very large number of different photometric measurements \citep{hoehn2014characteristics}. 



\section{Data}
\label{data}

The first dataset of galaxies used in this study contained galaxies from Sloan Digital Sky Survey \citep{york2000sloan} annotated manually by {\it Galaxy Zoo 2} \citep{willett2013galaxy} as galaxies that were not smooth and round (Question1 in {\it Galaxy Zoo 2} user interface). Since Galaxy Zoo is based on the annotations of non-experts, it cannot be assumed that all annotations are necessarily correct. To filter misclassified galaxies we used only galaxies on which 90\% or more of the voters agreed. That provided a dataset of 19,693 galaxies \citep{kuminski2014combining}, which are likely to be galaxies that are not smooth and round, and therefore are potentially spiral galaxies with identifiable handedness.

The ``superclean'' criterion \citep{lintott2011galaxy}, according which a classification is ``superclean'' only when it reaches 95\% of agreement, could not be used since using this criterion would leave merely 6,635 galaxies \citep{kuminski2014combining}, which might not be a sufficient number of galaxies for the analysis. However, the 90\% threshold is still higher than the ``clean'' {\it Galaxy Zoo} criterion \citep{lintott2011galaxy}, and therefore it can be reasonably assumed that the vast majority of the galaxies satisfying the criterion are indeed not smooth and round. Also, previous studies show that when using citizen science to classify between spiral and elliptical galaxies, the sensitivity of spiral galaxies is high, while the specificity of galaxies not annotated as spiral is lower \citep{dojcsak2014quantitative}, and therefore it is expected that the galaxies annotated as ``not smooth and round'' are indeed not elliptical galaxies.

The fact that all galaxies are bright and large allows correct classification of the galaxies regardless of their redshift, which in a randomly selected set of galaxies can lead to an inverse correlation between the classification accuracy and the z, regardless of whether the classification is carried out by a machine \citep{shamir2011ganalyzer} or by humans \citep{lintott2011galaxy}.

\begin{figure}[ht]
\plotone{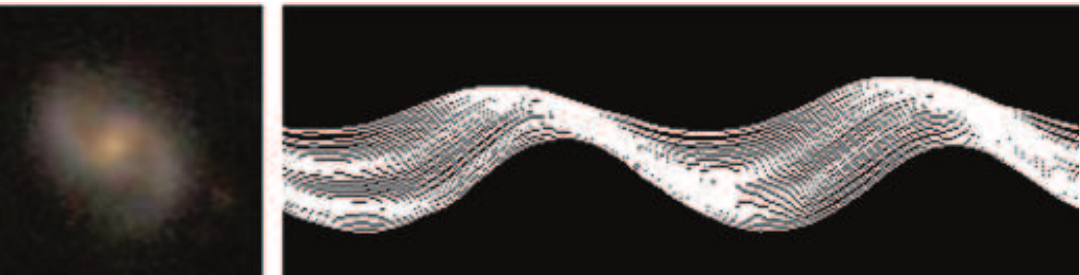}
\caption{A galaxy image and its corresponding radial intensity plot \citep{shamir2011ganalyzer}.}
\label{radial_intensity_plot}
\end{figure}

The galaxies were initially classified to clockwise and counterclockwise galaxies using the {\it Ganalyzer} galaxy image analysis tool  \citep{shamir2011ganalyzer,ganalyzer_ascl}. Ganalyzer works by first computing the Otsu binary threshold \citep{ots1979} to identify the foreground pixels, and the radius is determined by the most distant foreground pixel from the center of the galaxy. Then, the galaxy image is converted to its radial intensity plot, which is an image of 360$\times$35 pixels, such that the pixel $(x,y)$ in the radial intensity plot is the median value of the 5$\times$5 pixels around $(O_x+\sin(\theta) \cdot r,O_y-\cos(\theta)\cdot r)$ in the galaxy image, where ($O_x,O_y$) are the image coordinates of the galaxy center, $\theta$ is the polar angle (in degrees), and {\it r} is a radial distance, ranges over 35\% of the total galaxy radius. Figure~\ref{radial_intensity_plot} shows a galaxy image and its radial intensity plot.  
  
The horizontal lines in the radial intensity plot are searched for peaks, and the slope of the peaks determines the handedness of the galaxy. That is done by comparing each peak at coordinates $(x_0,y)$ to its closest peak in the next horizontal line $(x_1,y+1)$. If $x_1<x_0$ the counter {\it L} is incremented, and if $x_1>x_0$ the counter {\it R} is incremented. For the classification of the Galaxy Zoo data used in this study, if $L>R$ the galaxy is considered clockwise, and if $R>L$ it is considered counterclockwise. That is repeated for shifted radius ranges, from 20\%-55\% to 50\%-85\% of the total radius in increments of 10\%, until the sharpest slope is found. A detailed description about Ganalyzer can be found in \citep{shamir2011ganalyzer,shamir2012handedness,hoehn2014characteristics}. 

The automatic analysis described above is crude, and was therefore followed by manual inspection and correction of all galaxy classifications. The galaxies were then mirrored, and were inspected again to ensure that no galaxy is misclassified. The manual classification of the galaxies provided 6941 galaxies with clockwise pattern, and 6499 galaxies with counterclockwise pattern. The remaining galaxies did not have a clear identifiable handedness (e.g., edge-on), or were not spiral galaxies.

In the end of the process 100 galaxies from each class were selected randomly, and inspected carefully to ensure that all of them are correctly classified. The entire process of manual classification of the galaxies required approximately 150 hours of labor, but produced a very clean dataset of spiral galaxies separated by their spin direction, and no error in that dataset is believed to exist.

\begin{figure*}[ht]
\epsscale{2.4}
\plotone{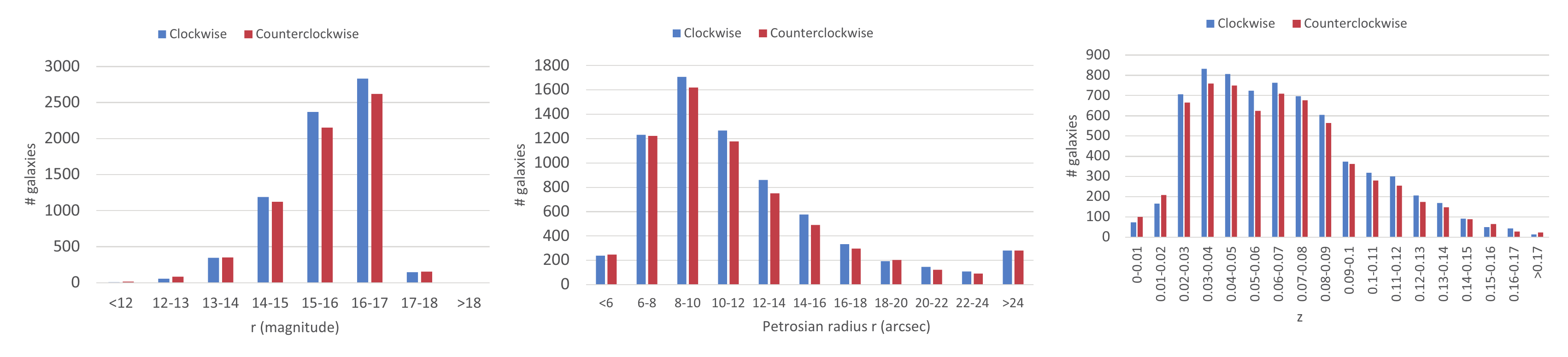}
\caption{Distribution of the r magnitude, Petrosian radius, and z of the galaxies in the dataset.}
\label{distribution}
\end{figure*}

The photometric information of each galaxy was retrieved through the Catalog Archive Server (CAS). All fields of the table PhotoObjAll in DR7 \citep{abazajian2009seventh} were used, producing a dataset of 452 variables for each galaxy. All galaxies were in the $90^o<RA<270^o$ hemisphere, and $z<0.25$. Figure~\ref{distribution} shows the histograms of the distribution of r magnitude, the Petrosian radius, and the redshift of the  galaxies with clockwise and counterclockwise patterns.  

\section{Classification method}
\label{method}

One of the goals of the experiment was to test whether the handedness of a spiral galaxy can be predicted using its photometric information. That was done by using several pattern recognition algorithms, such that the label of each galaxy sample is its handedness (cw or ccw), and the variables of each sample are the photometric variables from PhotoObjAll. The purpose of the supervised machine learning was to accurately predict the handedness of the galaxy using the photometric information.

For the classification, the Waikato Environment for Knowledge Analysis (WEKA) open source tool \citep{hall2009weka} was used. WEKA is a comprehensive software that includes the implementation of numerous  machine learning algorithms. The algorithms that were used for the classification were Random Forests \citep{breiman2001random}, OneR \citep{holte1993very}, Decision Table \citep{kohavi1995power}, Ensembles of balanced Nested Dichotomies \citep{dong2005ensembles}, Bayesian Network \citep{friedman1997bayesian}, and Bagging \citep{breiman1996bagging}. In addition to the algorithms provided by WEKA, the open source Weighted Nearest Distance (WND) algorithm \citep{shamir2008source,shamir2013wnd} was also used.

The experiments were performed such that 80\% of the samples were used for training, and the remaining 20\% were used for testing. That is, the machine learning algorithms used 80\% of the galaxies to automatically identify patterns that may differentiate between clockwise and counterclockwise galaxies, and the remaining 20\% of the galaxies were used to predict the handedness of each of these galaxies, and count the number of correct predictions such that the  classification accuracy was determined by the number of correct predictions divided by the total number of prediction attempts. 

In addition to the 80/20 strategy, the classification accuracy was also tested by separating the training and test data using a 10-fold cross-validation strategy, and also by using separation by fixed numbers such that 4800 samples from each class are used for training and 1200 samples from each class for testing.

The same experiments were also repeated such that the label (galaxy handedness) was replaced by a random label $\{0,1\}$. 

\section{Results}
\label{results}

\begin{figure}[p]
\epsscale{1.1}
\plotone{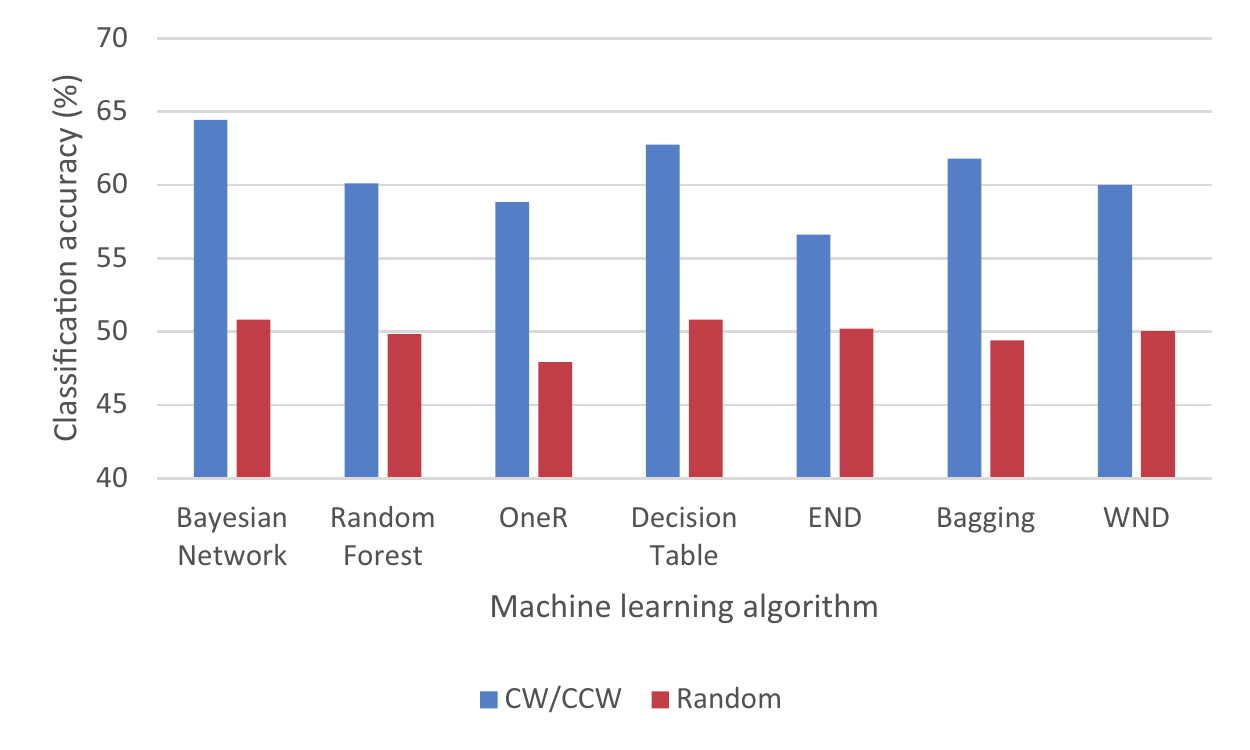}
\caption{Prediction accuracy of the galaxy handedness using SDSS photometric information. For each machine learning algorithm the graph shows the classification accuracy when the label is the correct handedness of each galaxy, and the accuracy when assigning random labels instead of the actual handedness.}
\label{exp_results}
\end{figure}

Figure~\ref{exp_results} shows the classification accuracy of the different supervised machine learning algorithms using labels that are the actual handedness  of each galaxy, as well as the randomly assigned labels.

Expectedly, not all algorithms achieved the same classification accuracy, as not all supervised machine learning algorithms are equally powerful, and different algorithms might perform differently on different types of data. However, it is clear that all algorithms classified the galaxies with accuracy higher than mere chance, while when the handedness was assigned randomly the classification accuracy of all algorithms was close to 50\%. When using fixed separation of 5000 samples per class for testing and 1000 per class for training the classification accuracy of the Bayesian Network marginally dropped to $\sim$63\%, and the classification accuracy of the WND algorithm dropped to $\sim$59\%.

Assuming no link between the variables and the galaxy handedness, a galaxy would be classified by these variables randomly to either clockwise or counterclockwise. In that case, achieving classification accuracy of 64.4\% by chance would require 1721 or more correct classifications of the 2688 total classification attempts. Using cumulative binomial probability \citep{keller2015statistics} such that the number of trails is 2688, minimum number of successes is 1721, and the probability of success is 0.5, the probability to have such results by mere chance is $<10^{-5}$. 

Tables~\ref{bayesnet} through~\ref{WND} show the confusion matrices of each of the classifiers. In all cases, both classes were classified in accuracy higher than mere chance.

\begin{table}[h]
\begin{center}
\caption{Confusion matrix of the classification with a Bayesian Network classifier}
\label{bayesnet}
{\small
\begin{tabular}{lcc}
\tableline\tableline
 & Clockwise & Counterclockwise \\
\tableline
Clockwise            & 888 & 480 \\
Counterclockwise & 487 & 833 \\
\tableline
\end{tabular}
}
\end{center}
\end{table}

\normalsize 

\begin{table}[h]
\begin{center}
\caption{Confusion matrix of the classification using Random Forests}
\label{random_forests}
{\small
\begin{tabular}{lcc}
\tableline\tableline
 & Clockwise & Counterclockwise \\
\tableline
Clockwise            & 948 & 420 \\
Counterclockwise & 652 & 668 \\
\tableline
\end{tabular}
}
\end{center}
\end{table}

\normalsize 

\begin{table}[h]
\begin{center}
\caption{Confusion matrix of the classification using OneR}
\label{OneR}
{\small
\begin{tabular}{lcc}
\tableline\tableline
 & Clockwise & Counterclockwise \\
\tableline
Clockwise            & 851 & 517 \\
Counterclockwise & 590 & 730 \\
\tableline
\end{tabular}
}
\end{center}
\end{table}

\normalsize 

\begin{table}[h]
\begin{center}
\caption{Confusion matrix of the classification using a Decision Table classifier}
\label{decision_table}
{\small
\begin{tabular}{lcc}
\tableline\tableline
 & Clockwise & Counterclockwise \\
\tableline
Clockwise            & 913 & 455 \\
Counterclockwise & 546 & 774 \\
\tableline
\end{tabular}
}
\end{center}
\end{table}

\normalsize 

\begin{table}[h]
\begin{center}
\caption{Confusion matrix of the classification using Ensembles of balanced Nested Dichotomies (END) classifier}
\label{END}
{\small
\begin{tabular}{lcc}
\tableline\tableline
 & Clockwise & Counterclockwise \\
\tableline
Clockwise            & 822 & 546 \\
Counterclockwise & 620 & 700 \\
\tableline
\end{tabular}
}
\end{center}
\end{table}

\normalsize 
 
\begin{table}[h]
\begin{center}
\caption{Confusion matrix of the classification using Bagging}
\label{bagging}
{\small
\begin{tabular}{lcc}
\tableline\tableline
 & Clockwise & Counterclockwise \\
\tableline
Clockwise            & 899 & 469 \\
Counterclockwise & 558 & 762 \\
\tableline
\end{tabular}
}
\end{center}
\end{table}

\normalsize 
  
\begin{table}[h]
\begin{center}
\caption{Confusion matrix of the classification using a Weighted Nearest Distance (WND) classifier}
\label{WND}
{\small
\begin{tabular}{lcc}
\tableline\tableline
 & Clockwise & Counterclockwise \\
\tableline
Clockwise            & 744 & 644 \\
Counterclockwise & 430 & 869 \\
\tableline
\end{tabular}
}
\end{center}
\end{table}

\normalsize 

\begin{figure}[p]
\plotone{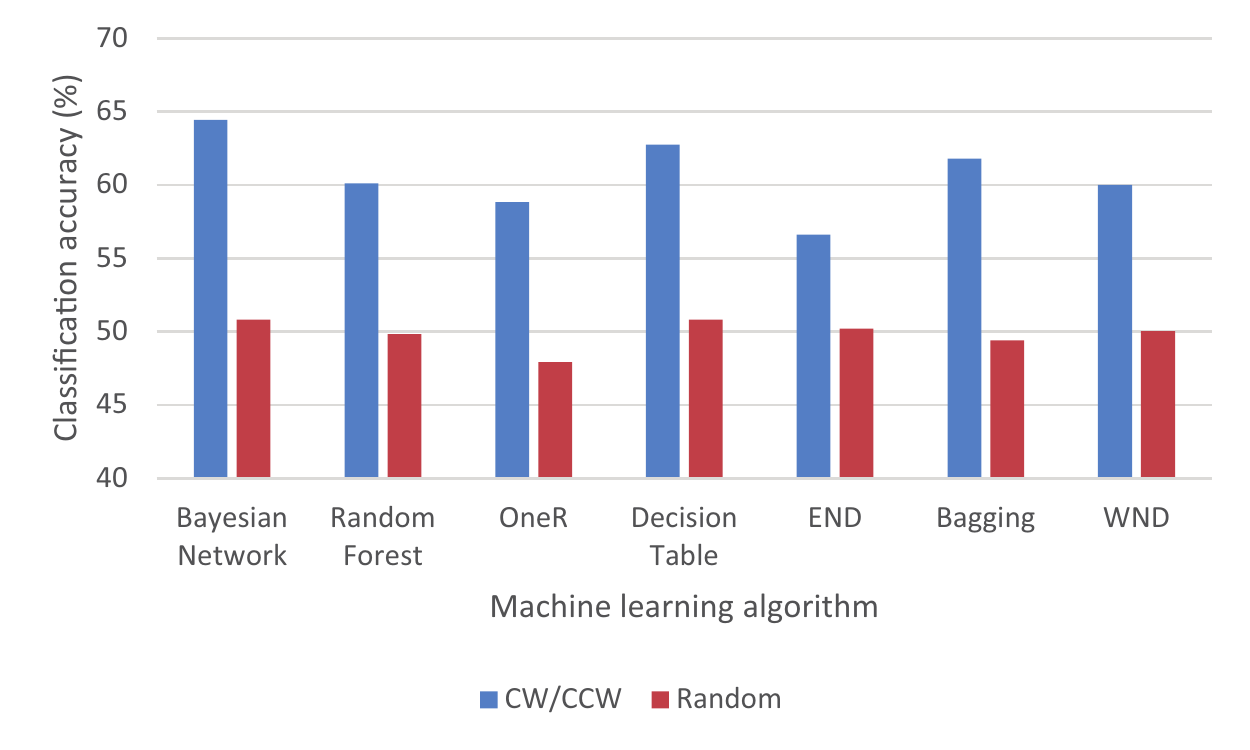}
\caption{Prediction accuracy of the spin direction when using 10-fold cross validation test strategy.}
\label{10fold}
\end{figure}

The classification accuracies of the different algorithms when using a 10-fold cross-validation strategy for testing are displayed by Figure~\ref{10fold}. Expectedly, the classification accuracies are similar to the classification accuracies when using 80\% of the samples for training. As with the 80/20 separation, when assigning the galaxies with random handedness the classification was close to 50\% mere chance accuracy. 

Because a large number of variables are being tested for the difference between clockwise and counterclockwise galaxies, the probability that one of these variables exhibits a difference by mere chance increases as the number of variables being tested gets larger. For that reason the Bonferroni correction is applied. Because the Bonferroni correction becomes stronger when the number of variables increases, using a large set of variables makes it less likely to identify specific variables that exhibit a Bonferroni-corrected statistically significant difference between the two classes of galaxies.

The Weighted Nearest Distance algorithm computes the Fisher discriminant \citep{fisher1938statistical} of each variable as a heuristics for determining the weight of each variable, such that higher weight indicates that the variable is assumed to be more informative for predicting the handedness of a spiral galaxy. The variables that were assigned with the highest Fisher discriminant scores are specified in Table~\ref{fisher_discriminant}. The table also displays the mean and standard error of the mean of the variables in clockwise galaxies and  counterclockwise galaxies, as well as the corrected and non-corrected two-tailed P value of the t-test of the difference between the means. Values such as -9999  and -1000 often appear in SDSS DR7 PhotoObjAll table, but these are in fact flags and not actual measured values, and therefore these values were ignored.

\begin{table*}[ht]
\begin{center}
\caption{Variables with the highest Fisher discriminant scores and t-test. No other variable showed a Bonferroni-corrected statistically significant difference between clockwise and counterclockwise galaxies.}
\label{fisher_discriminant}
{\scriptsize 
\begin{tabular}{lcccccc}
\tableline\tableline
Rank & Variable & Fisher          & Mean clockwise & Mean counterclockwise & t-test    & Bonferroni-corrected \\
        &             & discriminant  &                        &                                   &    P       & t-test P   \\   
\tableline
1 & isoPhiGrad\_r & 0.039 & 0.22$\pm$0.3 & 0.32$\pm$0.3 & 0.82 & 1 \\
2 & isoPhiGrad\_g & 0.026 & 0.051$\pm$0.33 & -0.004$\pm$0.33 & 0.13 & 1 \\
3 & isoPhiGrad\_i & 0.0076 & 0.05$\pm$0.33 & -0.21$\pm$0.32 & 0.58 & 1 \\
4 & petroR50Err\_u & 0.003 & 0.32$\pm$0.003 & 0.3$\pm$0.003  & 0.0004 & 0.16 \\
5 & lnLDeV\_i & 0.0027 & -3823$\pm$54.7 & -4187$\pm$65.5 & 0.00002 & 0.01 \\
6 & lnLStar\_u & 0.0027 & -1195$\pm$31 & -1400$\pm$41 & 0.00007 & 0.03 \\
7 & lnLDeV\_r & 0.0027 & -4885$\pm$66 & -5286$\pm$76 & 0.00007 & 0.03 \\
8 & lnLDeV\_z & 0.0026 & -690$\pm$17 & -800$\pm$20 & 0.00007 & 0.03 \\
9 & lnLDeV\_g & 0.0025 & -4993$\pm$69 & -5360$\pm$78 & 0.0004 & 0.18\\
10 & lnLDeV\_u & 0.0022 & -193$\pm$8 & -235$\pm$11 & 0.002 & 1 \\
11 & lnLStar\_z & 0.0022 & -9196$\pm$175 & -10313$\pm$214 & 0.00005 & 0.02 \\
12 & petroR90Err\_z & 0.002048 & -11.8$\pm$2.7 & -12.9$\pm$3 & 0.76 & 1 \\
13 & u\_g & 0.002 & -0.00057$\pm$0.0007 & 0.0032$\pm$0.0007 & 0.00006 & 0.02 \\
14 & u\_r & 0.0019 & -0.0016$\pm$0.0007 & 0.0022$\pm$0.0007 & 0.00007 & 0.03 \\
15 & lnLStar\_i & 0.002 & -38283$\pm$405 & -40384$\pm$450 & 0.0005 &  0.23 \\
16 & u\_i & 0.0019 & -0.0016$\pm$0.0007  & 0.0024$\pm$0.0007 & 0.00007 & 0.03 \\
17 & lnLStar\_g & 0.0018 & -34264$\pm$360  & -36000$\pm$400 & 0.001 & 0.63 \\
18 & u\_z & 0.0018 & -0.0012$\pm$0.0007  & 0.003$\pm$0.0007 & 0.00007 & 0.003 \\
\tableline
\end{tabular}
}
\end{center}
\end{table*}

\normalsize

When using the variables listed in Table~\ref{fisher_discriminant} only, the Bayesian Network classifier was able to differentiate between the two classes in accuracy of 63\%, and Random Forests and Bagging were able to achieve classification accuracies of 62\% and 61\%, respectively. That shows that this relatively small set of variables is sufficient to identify the handedness of the galaxies in accuracy higher than mere chance. 

The variables that have the highest Fisher discriminant scores are the isoPhiGrad\_r, isoPhiGrad\_g, and isoPhiGrad\_i, which are the isophote position angle gradients measured in the r, g, and i bands, respectively. Although their means did not exhibit a statistically significant difference between clockwise and counterclockwise spiral galaxies, the isophote position angle gradients were estimated to be the most informative by the Fisher discriminant heuristics. These variables are often used to measure ellipticity, and are not expected to be highly accurate (and for that reason were not included in the PhotoObjAll table of SDSS data releases after DR7). The isophotal position angle itself of clockwise and counterclockwise galaxies did not show any statistically significant difference.  

Of the full set of 452 variables, 10 showed a Bonferroni-corrected statistically significant difference between galaxies with clockwise pattern and galaxies with counterclockwise patterns. Out of these 10 variables, four are the SDSS `Stokes U' parameter u\_g, u\_r, u\_i, and u\_z, measured on the g, r, i, and z bands. The SDSS DR7 `Stokes U' parameter is measured by $U=\frac{a-b}{a+b}\sin(2\phi)$, where {\it a} is the major axis, {\it b} is the minor axis of the galaxy, and $\phi$ is the position angle \citep{abazajian2009seventh}. In all bands, the mean of the `Stokes U' parameter was negative for clockwise galaxies, and positive for counterclockwise galaxies.

The `Stokes U' parameter measured in the u band had a much lower non-corrected t-test statistical significance of $\sim$0.012, and therefore cannot be considered statistically significant. The SDSS `Stokes Q' parameter had no statistically significant difference in any of the bands, and its non-corrected t-test probabilities range between 0.31 for the r band and 0.96 for the u band. 



Other variables that show a statistically significant difference between clockwise and counterclockwise galaxies are lnLDeV and lnLStar, which provide information about the reliability of the separation of the object to stars and galaxies in the SDSS pipeline based on the magnitude model fitness. lnLDeV is the chi-square fitness of the de Vaucouleurs surface brightness model. The variables in the table related to that measurement are lnLDeV\_u, lnLDeV\_g, lnLDeV\_r, lnLDeV\_i, lnLDeV\_z, which measure the chi-square fitness of the de Vaucouleurs surface brightness model in bands u, g, r, i, and z, respectively. lnLStar measures the chi-square fitting of the PSF surface brightness model. The variables in Table~\ref{fisher_discriminant} related to lnLStar are lnLStar\_u, lnLStar\_g, lnLStar\_i, and lnLStar\_z, which measure the PSF model fitness in bands u, g, i, and z, respectively.


The remaining two variables in Table~\ref{fisher_discriminant} are petroR50Err\_u and petroR90Err\_z, which are the 50\% and 90\% Petrosian radius measurement error in the u and z bands, respectively. Consistent differences in these variables can be the results of weak measurements or differences in resolution, but can also be affected by different morphology. While these variables are assigned with relatively high Fisher discriminant scores and affected the classification, none of these variables show statistically significant difference between clockwise and counterclockwise galaxies.

Attempting to classify the galaxies with just the `Stokes U' parameter of the five bands provided very low classification accuracy of $\sim$50.8929 using a Bayesian Network classifier, indicating that while the differences based on galaxy handedness exist, they are not sufficient for predicting the handedness of a galaxy just by using that parameter. On the other hand, repeating the automatic classification experiments by removing all `Stokes parameters' did not make any significant impact on the classification accuracy, which remained as shown in Figure~\ref{exp_results}.

Since the galaxies that were used in the experiment are galaxies that were initially classified by crowdsourcing as spiral, the annotation of the galaxies can be subjected to human bias that in certain  conditions might affect the results. For instance, it has been shown that manual analysis of galaxy handedness by citizen scientists is substantially biased by human preferences \citep{land2008galaxy}. Although the handedness identification used in this study does not rely on citizen science annotations, it is possible that the human annotators have preferences to galaxies with certain combinations of handedness and other characteristics, and that bias might be carried forward to the dataset of galaxies separated by their handedness.

Such bias is expected to get weaker in galaxies on which more citizen scientists vote in the same manner, and therefore would exhibit itself in the form of smaller difference between the two classes when using galaxies on which the agreement among the citizen scientists was stronger. Table~\ref{superclean_fisher} shows the mean and standard error of the variables listed in Table~\ref{fisher_discriminant} measured using just 5132 of the galaxies that had clear handedness, and were also classified as spiral by 95\% or more of the citizen scientists.

\begin{table}[ht]
\begin{center}
\caption{Mean and standard error of the variables measured in galaxies which 95\% or more of the human annotators classified as spiral.}
\label{superclean_fisher}
{\scriptsize 
\begin{tabular}{lcc}
\tableline\tableline
Variable  & Mean            & Mean                                \\
              & clockwise      & counterclockwise               \\   
\tableline
isoPhiGrad\_r  & 0.58$\pm$0.46 & 0.48$\pm$0.49 \\
isoPhiGrad\_g  & 0.52$\pm$0.51 & -0.35$\pm$0.51  \\
isoPhiGrad\_i  & 0.69$\pm$0.49 & -0.206$\pm$0.322  \\
petroR50Err\_u  & 0.322$\pm$0.0058 & 0.3$\pm$0.0057  \\
lnLDeV\_i & -3855$\pm$88 & -4187.12$\pm$65.51 \\
lnLStar\_u  & -1218$\pm$54.2 & -1422.41$\pm$71.8  \\
lnLDeV\_r  & -4915$\pm$107 & -5356$\pm$125  \\
lnLDeV\_z  & -704$\pm$29.24 & -825$\pm$35.9  \\
lnLDeV\_g  & -5029$\pm$110 & -5485$\pm$132 \\
lnLDeV\_u  & -202$\pm$14 & -258$\pm$21  \\
lnLStar\_z  & -9320$\pm$280 & -10348.2$\pm$340.8  \\
petroR90Err\_z  & -8.5$\pm$1.7 & -8.25$\pm$1.1  \\
u\_g &  -0.00005$\pm$0.001 & 0.0034$\pm$0.0011  \\
u\_r & -0.0007$\pm$0.001 & 0.0024$\pm$0.0011  \\
lnLStar\_i  & -38777$\pm$642 & -40747$\pm$725  \\
u\_i & -0.0017$\pm$0.001  & 0.0024$\pm$0.0012  \\
lnLStar\_g  & -34667$\pm$572  & -36311$\pm$648  \\
u\_z & -0.00168$\pm$0.001  & 0.003$\pm$0.001  \\
\tableline
\end{tabular}
}
\end{center}
\end{table}

The differences between the means of the variables measured using galaxies classified as spiral by 95\% or more of the voters do not show a consistent increase compared to the differences when the entire dataset is used, indicating that the asymmetry does not change substantially with the voting trends of the citizen scientists. 

In addition to the Fisher discriminant feature selection, several other feature selection algorithms such as Consistency Subset Eval \citep{Liu1996}, Combined Feature Selection (CFS) Subset Eval \citep{Hall1998}, and Filtered Attribute Eval have also been used to automatically select the most informative variables, and the variables that were selected by these methods as well as the classification accuracy achieved using these variables are shown in Table~\ref{feature_selection}. As also mentioned above, these variables did not exhibit statistically significant difference between clockwise and counterclockwise galaxies. 

\begin{table}[ht]
\begin{center}
\caption{The variables selected by applying different automatic feature selection methods}
\label{feature_selection}
{ \scriptsize 
\begin{tabular}{lccc}
\tableline\tableline
Selection & Consistency    & Combined              & Filtered           \\
algorithm &                     &                             &                          \\   
\tableline
Selected  & petroR50Err\_u   & u\_r                 &  isoPhiGrad\_g      \\
Variables  &  petroR50Err\_g  &  isoPhiGrad\_u  &  isoPhiGrad\_r      \\
               &  petroR50Err\_r  &  isoPhiGrad\_g   &  isoPhiGrad\_i     \\
               &  petroR50Err\_i  &   isoPhiGrad\_r   &  isoPhiGrad\_u      \\
               &  petroR90Err\_i  &   isoPhiGrad\_i   &  isoPhiGrad\_z      \\
               &  petroR90Err\_z  &                        & petroR50Err\_i      \\
               &  u\_g                 &                      &  petroR50Err\_u      \\
               &  u\_r                  &                      &  petroR50Err\_g        \\
               &  u\_i                   &                      &  u\_r                        \\
               &  isoPhiGrad\_u     &                      &  petroR50Err\_r     \\
               &  isoPhiGrad\_g    &                      &   petroR90Err\_z     \\
               &  isoPhiGrad\_r     &                      &  petroR90Err\_i     \\
               &  isoPhiGrad\_i     &                      &  u\_i                      \\
               &  isoPhiGrad\_z    &                      &   u\_g                    \\
\tableline
Bayesian  & 63\%       &      62.9\%             &    63\%         \\
Network  &                    &                                 &                         \\
\tableline
Random &    62.8\%     &         61.9\%          &    62.8\%           \\
Forest  &                      &                                 &                          \\
\tableline
Bagging &    61.2\%      &        61\%              &     61.2\%        \\
\tableline
\end{tabular}
}
\end{center}
\end{table}

\normalsize

\subsection{Analysis using automatically classified galaxies}
\label{automatic}

The analysis using galaxies annotated by {\it Galaxy Zoo} involved a first step of manual classification of the galaxies to elliptical and spiral performed by crowdsourcing. The involvement of citizen scientists might therefore be affected by the human bias of the manual annotations. Although the classification by handedness was not performed by the citizen scientists, a certain preference of the human annotators to certain galaxies that depends on the galaxy handedness might be carried forward to produce a biased dataset.

To avoid such possible bias, another experiment was performed such that all galaxies were classified in a fully automatic manner, and without any human intervention in the analysis. That was done using galaxies from a computer-generated catalog of broad galaxy morphology \citep{kum16}. The catalog was generated using an automatic image classification method \citep{shamir2009automatic} applied to a large set of SDSS galaxies, producing a catalog of galaxies separated into elliptical and spiral galaxies as described in \citep{kum16}. That catalog is somewhat similar in its information to the {\it Galaxy Zoo 1} catalog, but was generated in a fully automatic manner, and without human intervention.

All galaxies with spectra classified as spirals were classified by the {\it Ganalyzer} algorithm \citep{shamir2011ganalyzer} described in Section~\ref{data} to determine their handedness. As discussed in Section~\ref{data}, the algorithm is imperfect and requires a step of intensive manual correction to produce a clean dataset. To avoid manual intervention, the algorithm was used such that the criteria for correct classification of a clockwise galaxy was the the {\it L} counter is 30 or higher, and $L>3R$. Similarly, the criteria for classifying a galaxy as counterclockwise is $R>30$ and $R>3L$. All other galaxies were considered undecided and were excluded from the analysis. That strategy improved the correctness of the classification to clockwise and counterclockwise galaxies, but also resulted in the sacrifice of 105,078 samples out of 115,359 galaxies. That provided a fully automatically-generated dataset of 5139 galaxies classified as clockwise, and 5142 galaxies classified as counterclockwise. Out of the dataset of 10,281 galaxies, 2280 overlap with the dataset described in Section~\ref{data} (1139 clockwise and 1141 counterclockwise). The distribution of the redshift, radius, and r magnitude of the galaxies are broadly consistent with those displayed in Figure~\ref{distribution}. 


To assess the consistency of the dataset, 200 galaxies from each class were inspected manually. The manual inspection revealed that out of the 400 galaxies, 11 galaxies classified as clockwise and eight galaxies classified by the algorithm as counterclockwise did not have clear handedness. One galaxy classified as counterclockwise was in fact a clockwise galaxy. 

\begin{figure}[p]
\epsscale{1.1}
\plotone{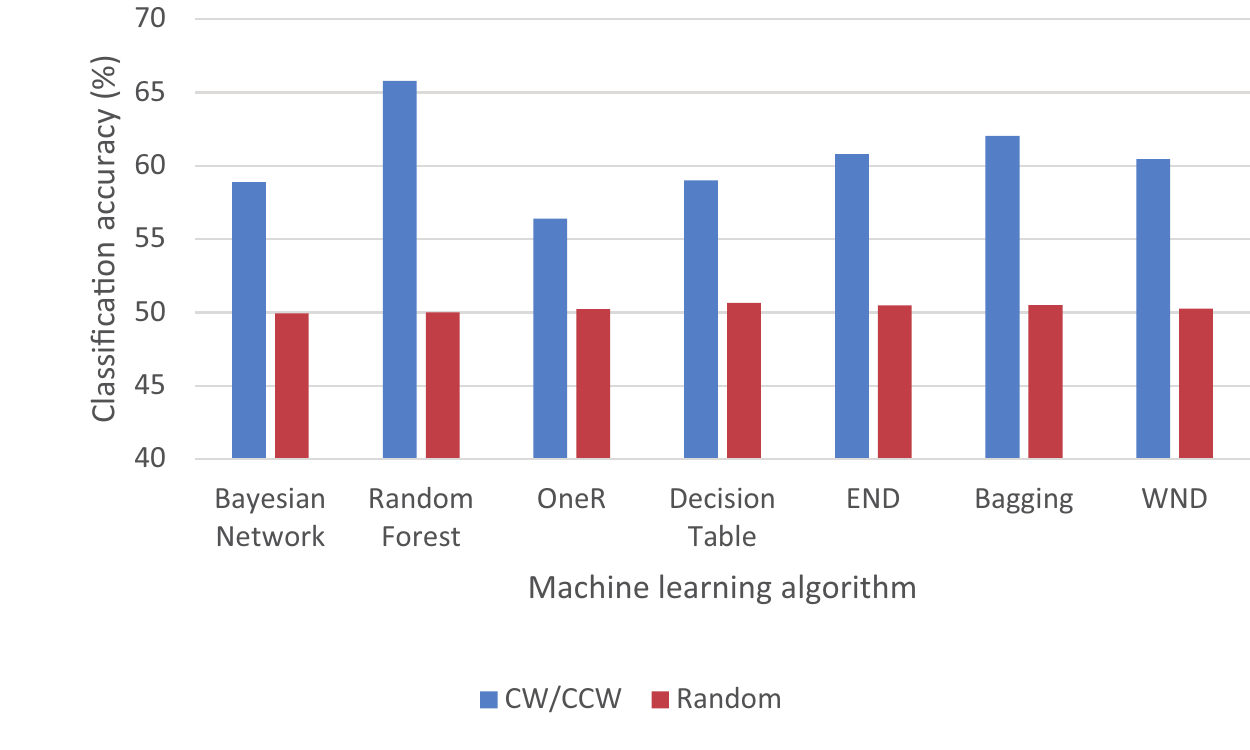}
\caption{Prediction accuracy of the galaxy handedness using SDSS photometric information of automatically classified galaxies.}
\label{automatic_results}
\end{figure}

The photometric information from SDSS DR7 for each galaxy was retrieved through CAS, and the dataset was classified similarly to the analysis of the Galaxy Zoo galaxies, with several different classifiers and a standard 10-fold test strategy. The classification results are displayed in Figure~\ref{automatic_results}.

As the figure shows, all classifiers were able to identify clockwise and counterclockwise galaxies with accuracy higher than mere chance, and the results are similar to the classification accuracy when the dataset was built based on the Galaxy Zoo galaxies. Tables~\ref{random_forests_automatic}, and~\ref{bagging_automatic} show the confusion matrices of the classifications using Random Forest and Bagging classifiers, respectively.

 \begin{table}[h]
\begin{center}
\caption{Confusion matrix of the classification using Random Forests}
\label{random_forests_automatic}
{\small
\begin{tabular}{lcc}
\tableline\tableline
 & Clockwise & Counterclockwise \\
\tableline
Clockwise            & 3480 & 1659 \\
Counterclockwise & 1858 & 3283 \\
\tableline
\end{tabular}
}
\end{center}
\end{table}

\normalsize 

\begin{table}[h]
\begin{center}
\caption{Confusion matrix of the classification using Bagging}
\label{bagging_automatic}
{\small 
\begin{tabular}{lcc}
\tableline\tableline
 & Clockwise & Counterclockwise \\
\tableline
Clockwise            & 3281 & 1858 \\
Counterclockwise & 1838 & 3304 \\
\tableline
\end{tabular}
}
\end{center}
\end{table}

\normalsize 

Like with the previous experiment using Galaxy Zoo galaxies, the informativeness of each variable was measured using Fisher discriminant heuristics, and the variables with the highest Fisher discriminant scores are listed in Table~\ref{fisher_discriminant_automatic}. 

\begin{table*}[ht]
\begin{center}
\caption{Variables with the highest Fisher discriminant scores}
\label{fisher_discriminant_automatic}
{\scriptsize 
\begin{tabular}{lcccccc}
\tableline\tableline
Rank & Variable & Fisher          & Mean clockwise & Mean counterclockwise & t-test    & Bonferroni-corrected \\
        &             & discriminant  &                        &                                   &    P       & t-test P   \\   
\tableline
1 & isoPhiGrad\_g & 0.027 & 0.41$\pm$0.36 & 0.16$\pm$0.36 & 0.6 & 1 \\
2 & isoPhiGrad\_r & 0.02 & -0.2$\pm$0.33 & 0.29$\pm$0.35 & 0.3 & 1 \\
3 & isoPhiGrad\_i & 0.004 & 0.68$\pm$0.37 & 0.34$\pm$0.37 & 0.5 & 1 \\
4 & deVPhi\_z & 0.002 & 92.1$\pm$0.74 & 88.7$\pm$0.74  & 0.001 & 0.477 \\
5 & deVPhi\_i & 0.002 & 91.4$\pm$0.75 & 88.5$\pm$0.75 & 0.006 & 1 \\
6 & mE1E2Err\_g & 0.002 & -1.08$\pm$0.48 & -0.43$\pm$0.3 & 0.25 & 1 \\
7 & deVPhi\_r & 0.002 & 91.4$\pm$0.76 & 88.1$\pm$0.75 & 0.002 & 0.689 \\
8 & expPhi\_i & 0.002 & 91.4$\pm$0.76 & 88.8$\pm$0.75 & 0.01 & 1 \\
9 & expPhi\_g & 0.002 & 90.6$\pm$0.76 & 87.7$\pm$0.75 & 0.007 & 1 \\
10 & expPhi\_z & 0.002 & 91.8$\pm$0.75 & 89$\pm$0.75 & 0.01 & 1 \\
11 & expPhi\_r & 0.002 & 91.1$\pm$0.76 & 88.45$\pm$0.75 & 0.01 & 1 \\
12 &  mE1E2Err\_r & 0.002 & -11.77$\pm$2.7 & -12.94$\pm$2.9 & 0.7668 & 1 \\
13 & deVPhi\_g & 0.002 & 90.6$\pm$0.76 & 87.9$\pm$0.75 & 0.01 & 1 \\
14 & mE1E2Err\_i & 0.002 & -1.08$\pm$0.48 & -0.43$\pm$0.3 & 0.25 & 1 \\
\tableline
\end{tabular}
}
\end{center}
\end{table*}

\normalsize 

Similarly to the first experiment, the variables with the highest Fisher discriminant scores were the isophote position angle gradients measured in the r, g, and i bands. Other variables are deVPhi\_z, deVPhi\_i, deVPhi\_r, and deVPhi\_g, which are the DeVaucouleurs fit position angles measured in bands z, i, r, and g, respectively, and the expPhi\_g, expPhi\_i, expPhi\_z, expPhi\_r, which are the exponential fit position angles measured in bands g, i, z, and r.  mE1E2Err\_g, mE1E2Err\_r, and mE1E2Err\_i are the square roots of the covariance matrix of the intensity second moments \citep{bernstein2002shapes}, measured on the g, r, and i bands.

The observation that the variables are associated with the position angle indicates on possible mild asymmetry between the morphology of these galaxies. Another explanation can be that inaccuracies in the measurement of the position angle might be sensitive to the handedness of the galaxy. That can also be related to the differences in the `Stokes U' parameter, which also depends on the position angle. However, since the position angles are randomly distributed, a consistent measurement error might lead to difference between the means measured in the two types of galaxies, but is not expected to result in classification between the types of galaxies based on the position angle alone. Also, when removing all variables related to the position angle (measured with the isophote, exponential fit, DeVaucouleurs fit, and `Stokes' parameters) and their errors, the classification accuracy is still higher than mere chance. The classification accuracy when removing these variables is $\sim$58\% and $\sim$57\% when using Random Forest and Bagging classifiers, respectively.  


Unlike the variables in Table~\ref{fisher_discriminant}, none of the variables in the PhotoObjAll table of DR7 showed statistically significant difference between clockwise and counterclockwise galaxies. That can be explained by the fact that the automatically generated dataset is not as clean as the dataset that was carefully inspected and corrected by manual intervention. Another possible explanation is that the human bias of the Galaxy Zoo citizen scientists was carried forward in some way, and possible preference of the human annotators of a certain handedness when attempting to differentiate between spiral and elliptical galaxies could be reflected by these variables. 

Table~\ref{feature_selection_automatic} shows the variables that were automatically selected by applying three different methods of automatic feature selection as was done in Table~\ref{feature_selection}, as well as the classification accuracy achieved when using these features alone. Most of these variables are the same variables identified by the Fisher discriminant heuristics listed in Table~\ref{fisher_discriminant_automatic}. The only exception is cx, which is the x of the unit vector of the right ascension and declination.

\begin{table}[ht]
\begin{center}
\caption{Variables selected by applying different automatic feature selection methods, and classification accuracy when using these features only.}
\label{feature_selection_automatic}
{ \scriptsize 
\begin{tabular}{lccc}
\tableline\tableline
Selection & Combined      & Consistency              & Filtered             \\
algorithm &                     &                             &                           \\   
\tableline
Selected  & isoPhiGrad\_u   &  isoPhiGrad\_u    &  isoPhiGrad\_g      \\
Variables  & isoPhiGrad\_g  &  isoPhiGrad\_g  &  isoPhiGrad\_u       \\
               &  isoPhiGrad\_r  &  isoPhiGrad\_r   &  isoPhiGrad\_r     \\
               &  cx                  &   isoPhiGrad\_i   &  deVPhi\_r           \\
               &                       &   deVPhi\_r        &  isoPhiGrad\_i   \\
               &                      &  expPhi\_g          & cx                      \\
               &                      &   expPhi\_r         &  expPhi\_g         \\
               &                     &   expPhi\_i           &  expPhi\_i           \\
               &                     &     cx                  &  expPhi\_r          \\
               &                     &                           &                         \\
\tableline
Bayesian  &     59.26\%   &          58.65\%        &      58.65\%       \\
Network  &                    &                              &                        \\
\tableline
Random &     61.33\%    &        65.73\%          &     65.73\%       \\
Forest  &                      &                             &                           \\
\tableline
Bagging &     59.93\%    &         63.25\%       &     63.25\%           \\
\tableline
\end{tabular}
}
\end{center}
\end{table}

\normalsize 

That table also shows that when using only the selected features the classification accuracy does not change substantially. When using all features except for the features selected by the Combined Feature Selection (CFS) algorithm the classification accuracy is 63.94\% with Random Forest and 65.48\% with Bagging. When removing the features selected by the Subset Attribute Eval and Filtered Attribute Eval the classification accuracy is 62.15\% and 68.04\% using Random Forest and Bagging classifiers, respectively. 

The experiment using the Galaxy Zoo galaxies showed that some of the magnitude model fitting variables showed statistically significant difference between clockwise and counterclockwise galaxies. These variables did not exhibit statistically significant difference when using the computer-generated dataset, but the classification accuracy when using the PSF  magnitude model fitting likelihoods (lnLStar), the exponential magnitude model (lnLExp) and the de Vaucouleurs magnitude model (lnLDeV) using all five bands the classification accuracy is higher than mere chance. Using Random Forest the classification accuracy is 56.1\%, and when using Bagging the classification accuracy is 54.5\%. When assigning the galaxies with random handedness, however, the classification accuracy is random. Table~\ref{magnitude_model} shows the confusion table of the classification when using the magnitude model variables and Random Forest classifier.

\begin{table}[h]
\begin{center}
\caption{Confusion table of the classification when using Random Forest classifier and only the 15 magnitude model fitting likelihoods variables lnLStar (PSF), lnLExp (exponential), and lnLDeV (de Vaucouleurs) measured in all five bands.}
\label{magnitude_model}
{\small
\begin{tabular}{lcc}
\tableline\tableline
 & Clockwise & Counterclockwise \\
\tableline
Clockwise            & 2998 & 2141 \\
Counterclockwise & 2376 & 2766 \\
\tableline
\end{tabular}
}
\end{center}
\end{table}

\normalsize

\section{Discussion}
\label{discussion}

Machine learning is typically used in astronomy for handling the vast pipelines of astronomical data \citep{fayyad1993skicat,djorgovski2006some}, and in particular analysis of galaxies \citep{ball2004galaxy,oyaizu2008galaxy}. In this study supervised machine learning was used to show differences between patterns of photometric variables of different types of galaxies -- galaxies with clockwise patterns and galaxies with counterclockwise patterns. In particular, statistically significant differences were observed in the SDSS ``Stokes U'' parameter, as well as the magnitude model fitting likelihoods.

The PSF fitting likelihood (lnLStar) variable is used in the SDSS pipeline to separate between stars and galaxy sources. When measured for galaxies, lnLStar can distinguish between flatter galaxies and galaxies that are less extended and are more point-like. The galaxies annotated by {\it Galaxy Zoo 2} have a relatively large surface size, but the higher PSF fitting likelihood shows that counterclockwise galaxies may be somewhat more dense, or have a more dominant nucleus, allowing a better PSF fitting.  

The de Vaucouleurs fit is often used to measure the variations in surface brightness of galaxies, and despite being normally used to profile elliptical galaxies, it can be considered more useful for profiling spiral galaxies compared to PSF fitting. The results show that counterclockwise galaxies, on average, exhibit a higher chi-square likelihood of fitting to the de Vacouleurs surface brightness distribution model. As mentioned above, the de Vaucouleurs surface brightness model was initially proposed for elliptical galaxies \citep{de1948recherches}, but in SDSS DR7 measurements were collected for all photometric objects. The higher likelihood of fitting the de Vaucouleurs surface brightness model might also suggest the existence a bright nucleus and a sharper and more consistent drop of brightness in spiral galaxies that rotate counterclockwise.

The experiment was performed with two different datasets of galaxies separated by their handedness. The first was based on galaxies classified as spiral by crowdsourcing, and the second was generated in a fully automatic process, and without the intervention of humans. The consistency between the experiments using the two different datasets indicates that the ability of a classifier to predict the handedness of a galaxy using its photometry data is not necessarily driven by bias of the manual galaxy annotations. 

The first dataset, in which spiral galaxies were selected by crowdsourcing, the number of galaxies with clockwise handedness was higher than the number of galaxies with counterclockwise handedness. That distribution of handedness disagrees with the handedness distribution in the dataset in which the galaxies were classified automatically, where the number of clockwise galaxies is slightly higher, but does not exhibit a statistically significant difference. That shows that the human classification of the spiral galaxies carried out by {\it Galaxy Zoo} was biased by the handedness, making a galaxy with clockwise handedness more likely to be voted as spiral compared to a galaxy with a counterclockwise pattern. These results show that studies carried out by applying the force of citizen scientists to analyze galaxy morphology can lead to biased results.

The asymmetry discussed here between clockwise and counterclockwise galaxies is observed through the galaxies classified by {\it Galaxy Zoo}, as well as dataset of galaxies classified in a fully automatic process. As the measurements for a large population of galaxies is expected to be symmetric, it is difficult to identify specific reasons for the ability of a classifier to predict the handedness of a galaxy based on its photometry. 

One possible explanation can be a consistent bias in the SDSS measurements taken form clockwise and counterclockwise galaxies. For instance, a consistent error in the measurement of the position angle that discriminates between clockwise and counterclockwise galaxies could affect several different measurements such as the SDSS `Stokes parameters'. Since in both datasets classification accuracy can be achieved by just using just a subset of the variables, such possible error possibly affects several different variables. Variables that allows the classification include variables affected by the position angle, but also variables that are not affected by the measurement of the position angle such as the magnitude model fitting likelihood variables. It will therefore require further investigation to profile the nature of the measurement bias and identify its source.

\section{Acknowledgments}

I would like to thank Dr. Noah Brosch and Dr. George Djorgovski for the helpful discussions and insightful comments that improved the paper. The research was supported in part by NSF grant IIS-1546079.

\appendix

\section{Appendix material}


The data files used in the experiments as well as computer-generated output files can be accessed at \url{http://vfacstaff.ltu.edu/lshamir/data/assym}. 

To perform the experiment using WEKA, the CSV file with the photometry information of the galaxies should be downloaded at \url{http://vfacstaff.ltu.edu/lshamir/data/assym/p_all_full.csv}, and opened using WEKA Explorer. In the ``classify'' tab, the field ``rotation'' should be selected as the class label, and then ``BayesNet'' (or any of the other classifiers) should be selected as the classifier. The results in this paper were produced by selecting percentage split of 80\% and 10-fold cross validation, but other test strategies can also be used.  

A file with randomized galaxy handedness can be downloaded at \url{http://vfacstaff.ltu.edu/lshamir/data/assym/p_all_full_randomized.csv}. 

The results using Weighted Nearest Distance can be produced using the input file \url{http://vfacstaff.ltu.edu/lshamir/data/assym/p_all_full_wndchrm.csv}. The CSV file can be analyzed directly using the UDAT software, which implements the WND classifier and can be downloaded at \url{http://vfacstaff.ltu.edu/lshamir/downloads/udat/}. After downloading the executable and the library files, the following command line should be used: \newline  

udat -w -r0.2 p\_all\_full\_wndchrm.csv wndchrm\_output.html  \newline     

The ``-w'' switch activates the Weighted Nearest Distance (WND) algorithm for classification, and ``-r0.2'' makes it randomly allocate 20\% of the samples for testing, and use the remaining 80\% for training.  

The resulting HTML file generated by the program can be viewed at \url{http://vfacstaff.ltu.edu/lshamir/data/assym/wndchrm_output.html}. The HTML file contains the classification accuracies, as well as other relevant information about the experiment such as the Fisher discriminant scores of the different variables that were used in the classification.  

Input file with randomized handedness can be downloaded at \url{http://vfacstaff.ltu.edu/lshamir/data/assym/p_all_full_wndchrm_randomized.csv}, and the results when using it can be viewed at \url{http://vfacstaff.ltu.edu/lshamir/data/assym/wndchrm_output_randomized.html}.  

For cross-validation, the following command line is used: \newline 

udat -w -r0.2 -n10 p\_all\_full\_wndchrm.csv wndchrm\_output.html \newline  

The resulting report file produced when using cross-validation can be viewed at \url{http://vfacstaff.ltu.edu/lshamir/data/assym/wndchrm_output_10fold.html}.

\bibliographystyle{apj}
\bibliography{galaxy_rotation_assym}

\clearpage



\clearpage


\end{document}